\documentclass[lettersize,journal]{IEEEtran}
\usepackage{amsmath,amsfonts}
\usepackage{algorithmic}
\usepackage{algorithm}
\usepackage{array}
\usepackage[caption=false,font=normalsize,labelfont=sf,textfont=sf]{subfig}
\usepackage{textcomp}
\usepackage{stfloats}
\usepackage{url}
\usepackage{verbatim}
\usepackage{graphicx}
\usepackage{cite}
\usepackage{xcolor}
\begin{document}

\title{Development of Silicon Micromachined Waveguide Filter-Banks for On-Chip Spectrometers}

\author{Matthew A. Koc, Jason Austermann, James Beall, Johannes Hubmayr, Joel N. Ullom, Michael Vissers, Jordan Wheeler
\thanks{Matthew A. Koc and Joel N. Ullom are with the Department of Physics at the University of Colorado, Boulder, CO, 80309, USA and the Quantum Sensors Division at the National Institute of Standards and Technology, Boulder, CO, 80305 USA. (email: matthew.koc@nist.gov)}
\thanks{Jason Austermann, James Beall, Johannes Hubmayr, Michael Vissers, and Jordan Wheeler are with the Quantum Sensors Division at the National Institute of Standards and Technology, Boulder, CO, 80305 USA}

}



\maketitle

\begin{abstract}
Development of high-speed, spatial-mapping spectrometers in the millimeter and far-infrared frequencies would enable entirely new research avenues in astronomy and cosmology. An “on-chip” spectrometer is one such technology that could enable Line Intensity Mapping. Recent work has shown the promise of high-speed imaging; however, a limiting factor is that many of these devices suffer from low optical efficiency. Here we present the fabrication of a metalized, Si waveguide filter-bank fabricated using deep reactive ion etching for use in millimeter spectroscopy. Our design simultaneously provides high-density pixel packing, high optical efficiency, high spectral resolution, and is readily compatible with simple and multiplexable MKID arrays. Gold plated test waveguide and filter show excellent match to simulations with a measured resolving power of 263 and a loss quality factor of 1116 at room temperature. The results show promise for extending the measurements to larger, multi-wavelength designs.
\end{abstract}


\section{Introduction}
The (sub)millimeter radiation bands are rich with cosmological and astronomical information. One emerging technique that could probe over 80\% of the volume of the observable universe is Line Intensity Mapping (LIM) \cite{Kovetz2019}. This technique images astrophysical emission lines as a function of redshift, which can provide deeper insight into areas of study such as dark matter and energy, inflation, and constraining $\Lambda$ cold dark matter ($\Lambda$CDM) cosmology \cite{karkare2022snowmass,choi2020atacama,hanson2013detection}. Each redshift bin provides a 2D map of matter distribution. One major benefit of LIM is that it can simultaneously measure a large number of bands across a wide band of frequencies of interest, in contrast to current design for most cosmic microwave background (CMB) receivers having comparatively fewer bands within the same bandwidth \cite{dutcher2023simons,niemack2010actpol,bhandarkar2025simons}.

\begin{figure}[tb]
\centerline{\includegraphics[width=\columnwidth]{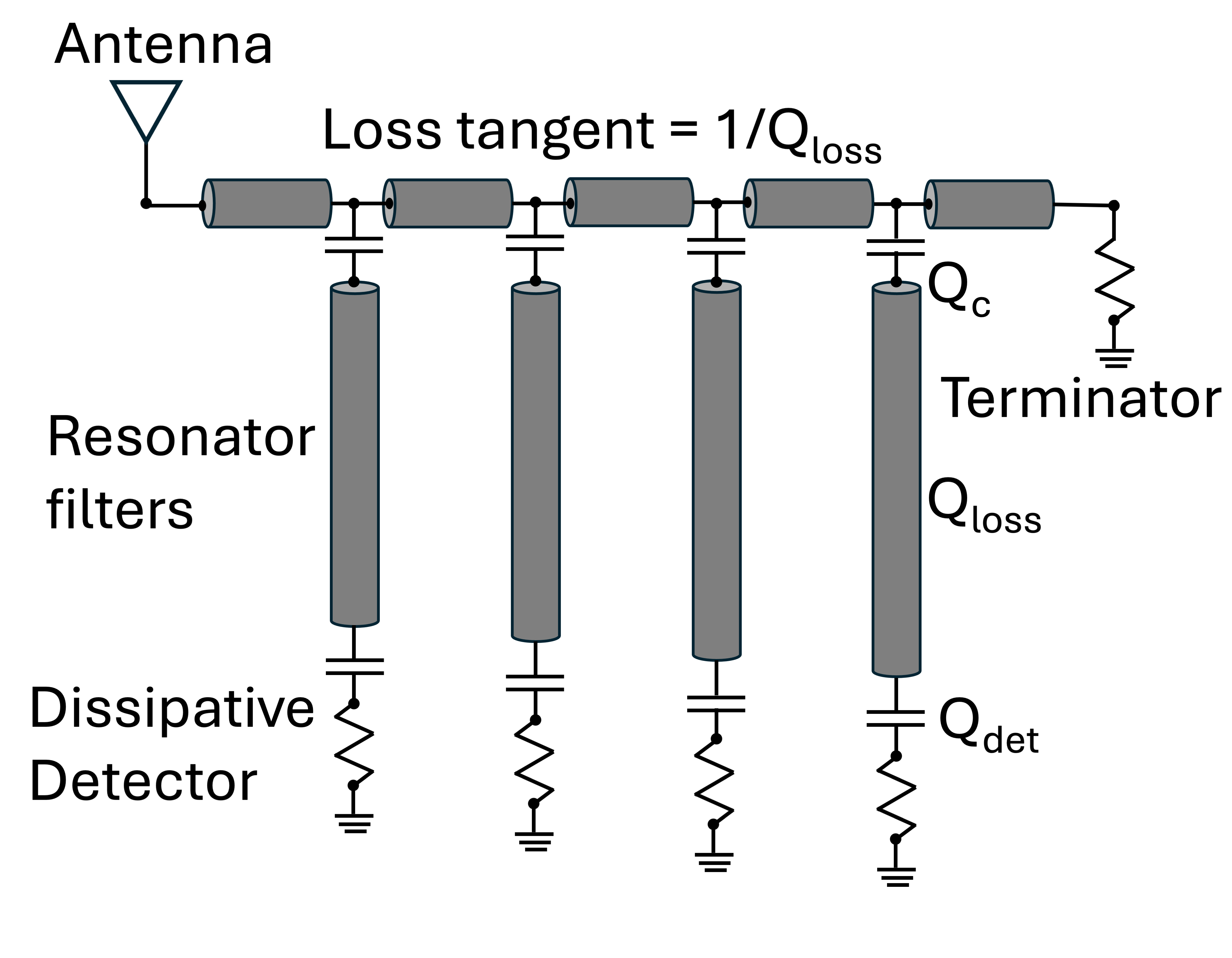}}
\caption{Filter-bank operation principle. Signal is coupled into a transmission line via an antenna. The power is coupled to resonant filters and then each filter couples to a detector. The terminator absorbs any power not absorbed or reflected by the filters, preventing any standing waves on the main transmission line.}
\label{figFBSchem}
\end{figure}

Multiple (sub)millimeter on-chip spectrometers have been proposed for LIM, including grating spectrometers \cite{cataldo2019MicroSPEC,switzer2021MicroSPEC,volpert2022MicroSPEC}, Fourier transform spectrometers \cite{basuthakur2020softs,basuthakur2022softs}, and Fabry-Perot interferometers \cite{nikola2022ccat,cothard2020design}. One promising technology for spectrally resolving (sub)millimeter light that has gained traction recently is the use of on-chip spectrometers using a filter-bank \cite{karkare2020SuperSPEC,redford2022,endo2019Deshima,taniguchi2022deshima2,bryan2016wspec,nie2024,Cecil2023sptslim_fab}. Filter-banks in the (sub)millimeter wavelength operate on the principles shown in Fig.~\ref{figFBSchem}. Light is coupled into the transmission line via an antenna. Light from the transmission line is then coupled to the spectral filters and then towards a detector with coupling strengths of $Q_c$ and $Q_{det}$, respectively. The maximum power that can be delivered to the detector occurs when $Q_c = Q_{det}$. Multiple resonant filters can be added in series to separate the light based on the frequency, creating a filter-bank spectrometer.  A single filter-bank spectrometer can fit within several cm\textsuperscript{2} of silicon, which approaches nearly an order of magnitude reduction in size compared to grating, Fourier transform, or Fabry-Perot spectrometers.

The maximum spectral efficiency of a filter-bank spectrometer, $\eta=50\%$, is given by
\begin{equation}
    \eta = \frac{1}{2} \left(1-\frac{R}{Q_{loss}}\right) \label{eta},
\end{equation}
where $Q_{loss}$ is the resonant quality factor describing any sources of loss in the filter-bank, and resolving power $R$, or quality factor of the filter-bank, is defined as 
\begin{equation}
    \frac{1}{R}= \frac{1}{Q_c} + \frac{1}{Q_{det}} + \frac{1}{Q_{loss}} \label{R}.
\end{equation}
Here $Q_c$ is the coupling quality factor between the feed-line and the filter and $Q_{det}$ is the coupling quality factor between the filter and the detector \cite{robson2022}.  These equations demonstrate the crucial need for low-loss (high $Q_{loss}$) transmission lines as that parameter limits both the resolving power and the spectral channel efficiency. Although (sub)millimeter on-chip spectrometers have been under development for over a decade \cite{Shirokoff2012,Endo2012}, one major limitation has been the efficient transmission of photons from coupling optics to spectroscopic detection. The highest achieved optical efficiency reported is approximately a factor of 3 lower than theoretical maximum, indicating there is room for improvement in design and development \cite{endo2019Deshima, redford2022}. 

We propose a metalized, micromachined Si filter-bank waveguide designed in a split-block architecture. Because the transmission media of the filter-bank is vacuum, these waveguides represent the lowest potential loss of all materials; however, there is still the potential of loss through the side walls due to finite conductivity and surface roughness.  

\section{Methods}

\begin{figure}[tb]
\centerline{\includegraphics[width=\columnwidth]{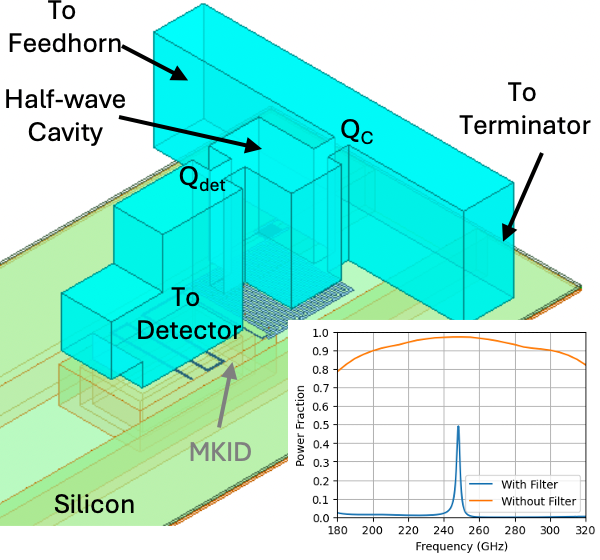}}
\caption{The 3D realization of a single channel filter-bank as a high-frequency structure simulator (HFSS, Ansys, see Section~\ref{footnote}) model. The waveguide transmission line couples to a half-wave resonant filter, which in turn couples to a waveguide with an H-plane bend terminating into an MKID on a separate wafer. The inset shows the detector response with and without the filter. With the filter nearly perfect, 50\% coupling can be achieved. This detector is on a 10 $\mu$m SOI membrane with a separate deep-etched metalized wafer serving as a quarter-wave backshort.  }
\label{figdetector}
\end{figure}

\subsection{Design}
\begin{figure}[tb]
\centerline{\includegraphics[width=\columnwidth]{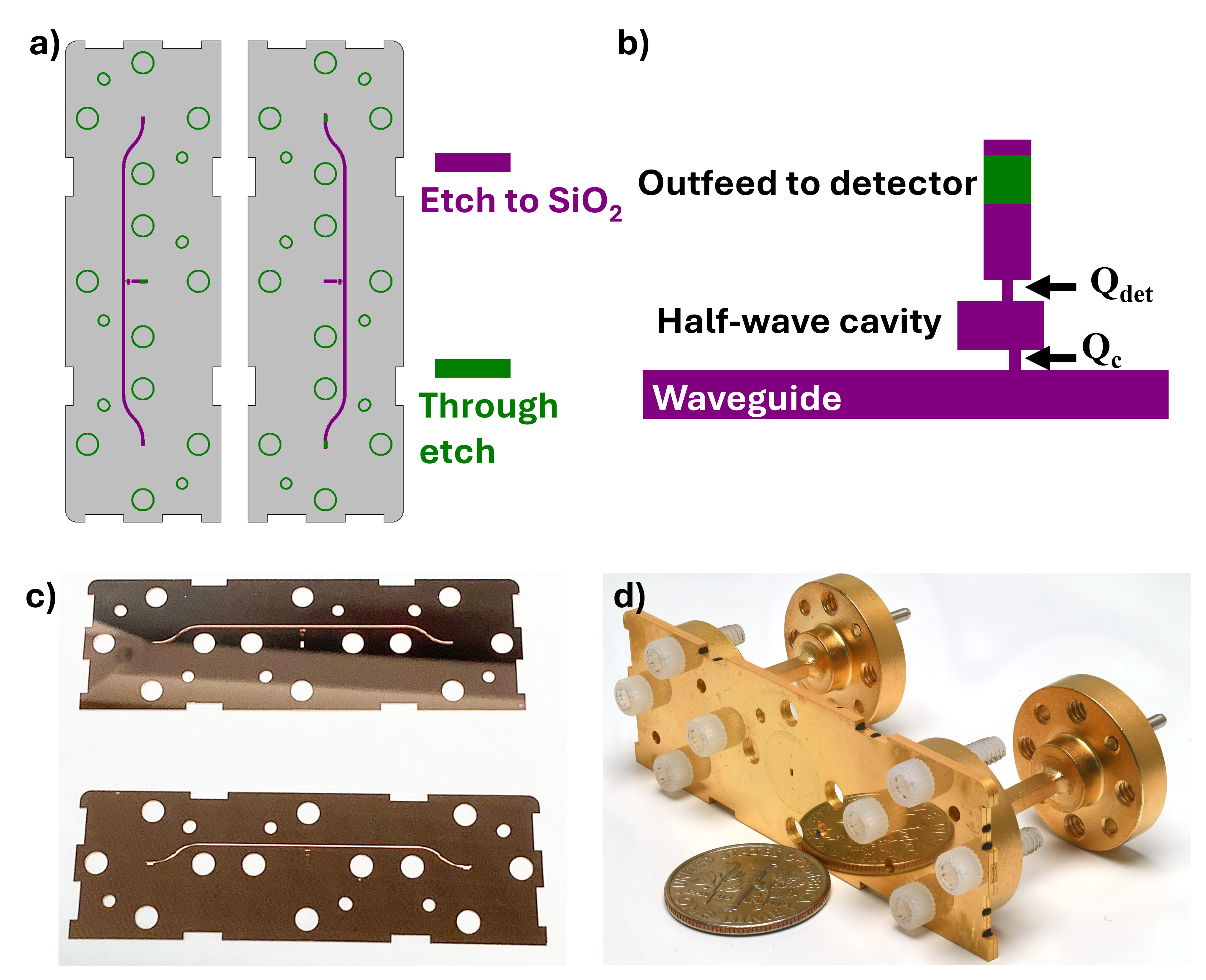}}
\caption{Designed and fabricated test parts of a single filter waveguide test part. a) shows the full split-block design where the purple traces show an etch that stops on the SiO\textsubscript{2} layer and the green is etched completely through the wafer. Each part of the split-block is almost a perfect mirror, but the feedhorn, terminator, and detector ports are not symmetric between the two halves. b) shows the general shape of the filter. The wavelength is determined by the length of the half-wave cavity and quality factors are set by the position and width of the coupling slots. c) shows the fabricated device after an initial Ti/Cu seed layer. d) shows the assembled, Au-plated waveguide with attached commercial VNA coupling parts.}
\label{figDesign}
\end{figure}

Here we present a single-channel waveguide filter-bank as a proof of concept demonstration. The waveguide architecture shown in Fig.~\ref{figdetector} can achieve a theoretical maximum $\eta$=50\% for a single filter when $Q_c=Q_{det}$, based on standard transmission line circuit analysis. The half-wavelength waveguide cavity acts as the resonant filter, while the coupling strengths $Q_c$ and $Q_{det}$ are controlled by capacitive slots. An H-plane 90\textdegree ~bend brings the waveguide channel outputs to a detector. To realize this geometry, we create an E-plane split block waveguide by micromachining mirrored parts, as shown in Fig.~\ref{figDesign}a. This geometry minimizes any loss due to reflections and radiation through small gaps because the currents in the TE$_{10}$ mode do not cross the split plane. Images of a functioning, realized single-channel filter-bank are shown in Fig.~\ref{figDesign}c and Fig.~\ref{figDesign}d.

\subsection{Fabrication}
Micromachined parts were fabricated using a 150 mm silicon-on-insulator (SOI) wafer with a 430 $\mu$m handle thickness, 0.5 $\mu$m oxide thickness, and 220 $\mu$m device side thickness. SOI wafers were used to ensure a consistent etch depth across the entire wafer. The test device was etched in 3 steps: primary alignment marks, back side deep etch and front side deep etch. The resulting micromachined parts were aggressively cleaned to remove any residual etch byproduct on the side walls and were then metallized. All lithography was performed using direct-write photolithography.

\subsubsection{Primary alignment marks}
Alignment marks were lithographically placed in regions that would not affect the waveguide properties. Alignment marks were patterned into SPR 660 photoresist (Dow, see Section~\ref{footnote}) on the device-side layer. Marks were etched with a SF\textsubscript{6} plasma etch process in a Deep Reactive Ion Etcher (DRIE) resulting in $\sim$1 $\mu$m etch depth.

\subsubsection{Back side deep etch}
The back side deep etch was patterned onto the handle side of the wafer with SPR 220-7 photoresist (Dow, see Section~\ref{footnote}) spun at approximately 7 $\mu$m thickness. The primary alignment marks on the opposite side of this pattern were used to align the pattern in the maskless aligner. The back side deep etch pattern includes all features intended to be etched completely through the wafer, such as the perimeter, screw holes, and entrance/exit ports for coupling light into/out of the waveguide spectrometer. 

The features were etched down 430 $\mu$m to the oxide layer with a DRIE using a modified Bosch etching process in an SPTS Rapier (see Section~\ref{footnote}). Briefly, the etch process involved three steps which were looped sequentially until the etch depth was reached. First a C\textsubscript{4}F\textsubscript{8} plasma was used to deposit an isotropic, passivating CF\textsubscript{x} film to protect the side walls and cover the exposed Si, followed by a directional O\textsubscript{2} etch to clear the bottom passivation film, finally followed by an isotropic SF\textsubscript{6} etch to rapidly remove exposed Si. The platen temperature was set to 0 \textdegree C for this process. The endpoint of the etch was monitored by an optical endpoint detector which followed the intensity of the Si-F optical emission line at 440 nm \cite{armstrong1979spectroscopic}. The SiO\textsubscript{2} layer is only weakly reactive using this process, so it acts as an  etch stop, allowing for a uniform etch depth across the wafer. Once the SiO\textsubscript{2} layer is exposed, the concentration of SiF\textsubscript{4} in the plasma decreases which can easily be measured in the optical emission spectrum. After visual confirmation of completely etching to the oxide layer across the entire device, the oxide is removed using a low power CF\textsubscript{4}/O\textsubscript{2} reactive ion etch.

\subsubsection{front side deep etch}
The etched back side of the wafer was laminated with roll-on negative photoresist with a BoPET backing to protect the electrostatic chuck in the DRIE once the wafer was etched through. The front side was patterned with $\sim$7 $\mu$m photoresist as described in the back side deep etch. The features were etched using the same Bosch process on the DRIE and the etch was monitored with the optical emission end point detector as described above. This step etches the waveguide and filter-bank features at a depth of 220$\mu$m, due to the device-side thickness. It also simultaneously etches through the remaining silicon on the through-etch portions, such as the device perimeter and screw holes. Accurately stopping the etch once the endpoint is reached is critical, because etching longer can result in more severe undercut, altering the frequency set by the dimensions of the half-wavelength cavity. The 0.5 $\mu$m oxide layer in the waveguide and filter-bank features was cleared using the CF\textsubscript{4}/O\textsubscript{2} reactive ion etch described above.

\subsection{Preparation for metallization}
The etch byproduct on the sidewalls from deep etching is notorious for preventing conformal metallization \cite{britton2010corrugated,nibarger2012}. To address this, each part was cleaned through an aggressive cleaning process. Photoresist was removed by soaking in a bath of acetone for 3 min, sonicating in a clean bath of acetone for 3 min, then rinsing in a bath of IPA for 3 min. The parts were allowed to dry overnight. Samples were then cleaned in 75 \textdegree C Nano-strip (FUJIFILM Electronic Materials, see Section~\ref{footnote}) for 15 min, then were rinsed in a quick dump rinse (QDR) tank with DI water for 20 cycles. The parts were then cleaned in a bath of 55 \textdegree C Novec7200 (3M, see Section~\ref{footnote}) for 15 min and a bath of 70 \textdegree C PlasmaSolv EKC265 (DuPont, see Section~\ref{footnote}) for 15 min. Residual EKC 265 was removed from the parts with 2 baths of IPA for 3 min, 20 cycles of QDR, then a final bath of IPA for 3 min. Parts were allowed to air dry undisturbed overnight.

\subsection{Metallization and assembly}
Multiple, different metallization methods were attempted; however, only electroplating with Cu and Au yielded a functional device. The Au plating was performed in a 2 step process at external vendors. A seed layer was performed at LGA Thin Films, Inc. (Santa Clara, CA, USA, see Section~\ref{footnote}) 
where parts underwent an in situ sputter clean followed by 200 nm Ti topped with 1 $\mu$m of Cu at high pressure. Parts were then electroplated at Custom Microwave Inc. (Longmont, CO, USA, see Section~\ref{footnote}) resulting in 3 $\mu$m of electroplated Cu followed by 3 $\mu$m of Au. 

Test devices were also coated with sputtered Al in-house. Cleaned Si parts were mounted to a carrier wafer. Al was deposited via sputtering in an Ar plasma with the carrier wafer cooled to 25 \textdegree C and rotated at 40 rpm to ensure a uniform coating. Multiple deposition times were tested from 20 min to 40 min per side.

The two mirrored parts were assembled on granite blocks, aligned with vernier marks \cite{reck2013measurement} to within 2 $\mu$m on all four edges. Multiple screws with Teflon washers were used to temporarily clamp the parts together while STYCAST 2850FT (Henkel Loctite, see Section~\ref{footnote}) was applied along the edges to permanently hold the parts together.

\section{Results and Discussion}

\begin{figure}[tb]
\centerline{\includegraphics[width=\columnwidth]{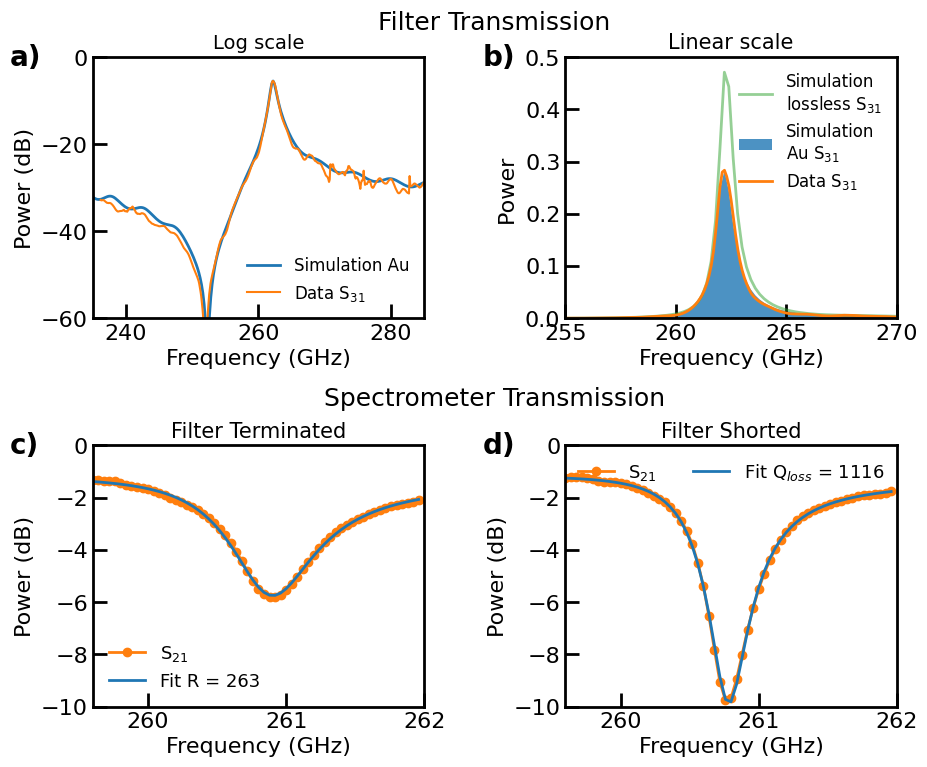}}
\caption{Experimental and simulated results for the Au-plated waveguide filter in the log (a) and linear (b) scales. The Au-plated simulation (blue) matches the measured data (orange). The lossless waveguide was simulated (green) to show a maximum expected efficiency for this design. (c) measured power loss of the filter when it is terminated resulting in a fit of $R=263$. (d) measured power loss when the filter is shorted resulting in a fit of $Q_{loss}=1116$.}
\label{figAuWG}
\end{figure}

A single-channel (260 GHz) Au-plated waveguide filter-bank was fabricated and tested at room temperature using a mm-wave vector network analyzer (VNA). The VNA experimental setup has been previously presented \cite{britton2010corrugated}. Commercial VNA adapter components were used to test the design measuring the transmitted frequency through the filter and the spectrometer as seen in Fig.~\ref{figAuWG}. The data show a good match to HFSS (Ansys, see Section~\ref{footnote}) simulation of Au-plated waveguide with a slight shift (0.5\%) in frequency. 
We measure a filter efficiency of nearly 30\%, which matches the Au simulation. We find that $S_{11}$ is less than -10 dB, indicating that the H-plane 90\textdegree ~bend does not have a significant reflection. The resolving power is fit to be $R=263$ (designed $\sim$ 300), as shown in Fig. ~\ref{figAuWG}c. 

By terminating and shorting the filter output (effectively setting $Q_{det} \gg Q_{loss}$) and measuring the transmitted power through the output of the spectrometer, we determine $Q_{loss} = 1116$, as shown in Fig.~\ref{figAuWG}d. We can calculate our maximum spectral efficiency based on \eqref{eta} to be $\eta=0.384$ at room temperature. Because the $Q_{loss}$ can improve as the square root of the residual-resistance ratio (RRR), it's reasonable that a significant improvement in loss can be achieved upon cooling. However, the anomalous skin effect will limit the high-frequency loss reduction when compared to the DC loss dictated by RRR alone \cite{Chambers1952}. 

\begin{figure}[tb]
\centerline{\includegraphics[width=\columnwidth]{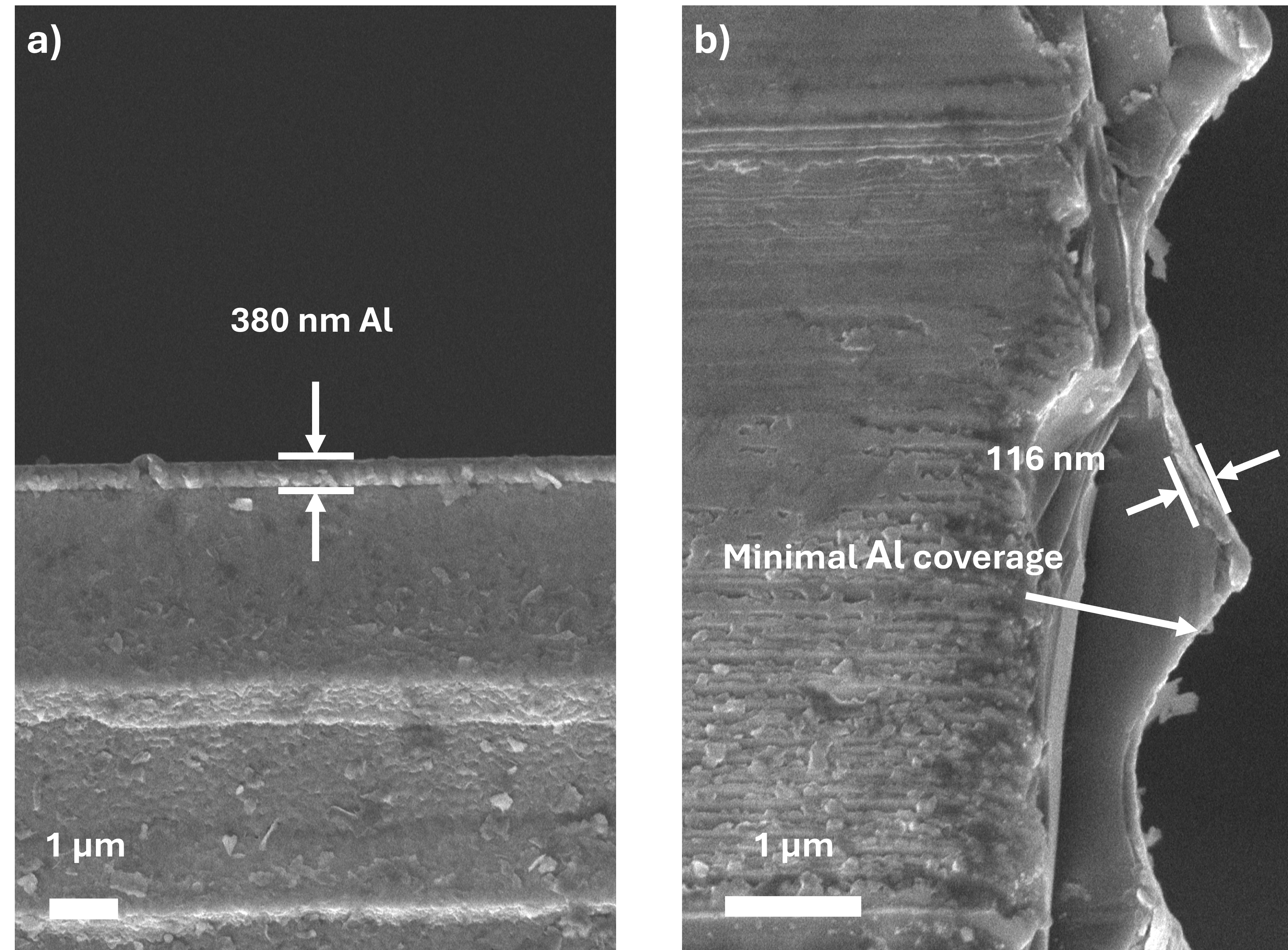}}
\caption{SEM images of DRIE etched waveguide parts, sputtered with Al. For a process that results in 380 nm of Al deposited on the top surface (a) only a fraction can coat the side walls within the etch cavity (b). The upward facing surface has $\sim$100 nm of Al while the downward, shadowed surface has minimal Al coverage. Resultant devices metallized in this manner showed excessive loss.}
\label{figSEM}
\end{figure}

Waveguides metalized using only sputtered metal resulted in extremely lossy waveguides. Because the Bosch etching process results in a sidewall that is scalloped, even with a high-pressure sputter, the underside of the scallop is shadowed and little-to no metal is adhered, as seen in Fig.~\ref{figSEM}, resulting in a waveguide with lossy sidewalls. Thus, the need for mass transport within the cavities during the metallization step is indicated and has been addressed by electroplating. Lossy waveguides were observed with multiple conditions of in-house Al sputtering as well as parts with the Ti/Cu sputtered seed layer performed by the external vendor. 

\begin{figure}[tb]
\centerline{\includegraphics[width=\columnwidth]{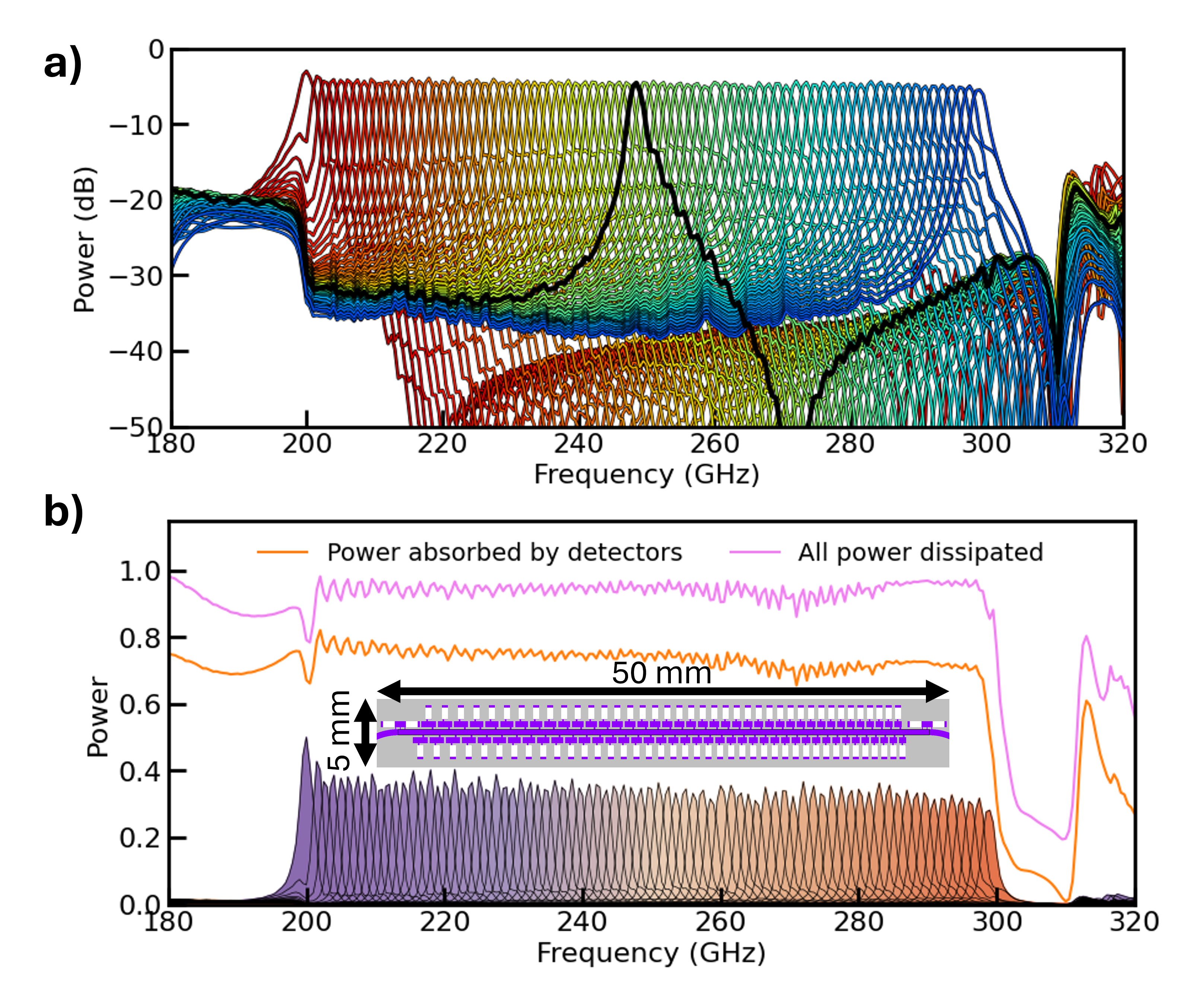}}
\caption{Simulation of a full filter-bank spectrometer. The simulation is made via cascading all of the S matrices of individual filters together and each filter to a detector simulation S matrix. A filter-bank was designed with 80, $R=100$ channels with 1.5 bandwidth. a) shows the spectrometer performance for each individual filter using a log scale. One specific filter is highlighted (black) to show the general shape of each filter. Out of band coupling withing the spectrometer operation range of 200 GHz to 300 GHz is less than 30 dB. b) shows filter response in linear scale and also the the cumulative power absorbed by the detectors (orange) and the total dissipated power (purple). 
The difference between the orange and purple curves is due to dissipation from $Q_{loss}$.The inset shows a schematic of the spectrometer size.}
\label{figFullSpec}
\end{figure}

Given the achieved $Q_{loss}$ in excess of 1000 it is reasonable to build spectrometers with resolving powers of 100 or less without further development to improve loss. As such, the simulated performance of a full spectrometer is shown in Fig.~\ref{figFullSpec}. The spectrometer is composed of 80 channels with $R=100$ and a bandwidth of 1.5. The overall dimension of the filter-bank is 5$\times$50 mm\textsuperscript{2}, so an array of 20 of these filter-banks could easily be fabricated using a single 150 mm SOI wafer and coupled to a single monolithic 1,600 detector MKID array. This spectrometer card could then be oriented in the z-direction of a sub-millimeter focal plane so that it only occupies 5$\times$100 mm\textsuperscript{2} of focal plane real estate. A stack of 20 spectrometer cards would make a 400-pixel, 32,000 detector spectrometer with a focal plane footprint of only $\sim$100$\times$100 mm\textsuperscript{2}. 
This spectrometer is designed to efficiently transmit light in the range of 200 GHz to 300 GHz, which would be capable of measuring the CO and CII emission lines \cite{karkare2022snowmass}. Presuming a similar efficiency to the single-filter that we measured, we expect $\sim$74\% efficiency across the band. For comparison, a nearly lossless filter-bank could result in up to 90\% efficiency across the frequency range. 

\section{Conclusion}
The Au-plated, single filter results demonstrate one of the highest reported filter efficiencies to date. This technology is highly scalable and will enable accurate spectroscopic measurements with high-spatial mapping speeds. By utilizing an E-plane split-block construction we can minimize any losses due to gaps in assembly, resulting in a highly efficient spectrometer. A further increase in efficiency should be possible by metallizing with a superconducting metal, such as Nb. We were unsuccessful in complete sidewall coverage when exclusively using sputtering as the technique for metallization and require a method such as high-pressure seed layer sputtering or Atomic Layer Deposition followed by electroplating to ensure complete coverage.

\section{Notes}\label{footnote}
Products or companies named here are cited only in the interest of complete technical description, and neither constitute nor imply endorsement by NIST or by the US government. Other products and companies may be found to serve just as well.

\section*{Acknowledgments}
Thanks to Jeff Van Lanen for assistance in aligning and gluing the plated waveguide test parts together.

 
%

\bibliographystyle{IEEEtran}
\bibliography{IEEEabrv,mak_LTD}

\vfill

\end{document}